\documentclass[a4paper]{jpconf}
\usepackage{graphicx}
\begin{document}
\title{The effect of $^{12}$C + $^{12}$C rate uncertainties on s-process yields}

\author{M E Bennett$^1$, R Hirschi$^{1,2}$, M Pignatari$^{7,3,4}$, S Diehl$^5$, C Fryer$^6$, F Herwig$^7$, A Hungerford$^6$, G Magkotsios$^{3,8}$, G Rockefeller$^6$, F Timmes$^8$, M Wiescher$^3$ and P Young$^8$.}

\address{$^1$ Astrophysics Group, Keele University, ST5 5BG, UK}
\address{$^2$ IPMU, University of Tokyo, Kashiwa, Chiba 277-8582, Japan}
\address{$^3$ Joint Institute for Nuclear Astrophysics, University of Notre Dame, IN, 46556, USA}
\address{$^4$ TRIUMF, 4004 Wesbrook Mall, Vancouver, BC, Canada, V6T 2A3}
\address{$^5$ Theoretical Astrophysics Group (T-6), LANL, Los Alamos, NM, 87544, USA}
\address{$^6$ Computation, Physics and Methods (CCS2), LANL, Los Alamos, NM, 87544, USA}
\address{$^7$ Dept. of Physics \& Astronomy, Victoria, BC, V8W 3P6,Canada}
\address{$^8$ School of Earth and Space Exploration, University of Arizona, Tempe, AZ 85287, USA}

\ead{meb@astro.keele.ac.uk}

\begin{abstract}
The slow neutron capture process in massive stars (the weak s-process) produces most of the s-only isotopes in the mass region $60 < A < 90$.  The nuclear reaction rates used in simulations of this process have a profound effect on the final s-process yields.  We generated 1D stellar models of a 25$M_{\odot}$ star varying the $^{12}$C + $^{12}$C rate by a factor of 10 and calculated full nucleosynthesis using the post-processing code PPN.  Increasing or decreasing the rate by a factor of 10 affects the convective history and nucleosynthesis, and consequently the final yields.
\end{abstract}

\section{Introduction}

Elements in the solar system are formed from a variety of nucleosynthesis processes in stars, such as the slow and rapid neutron-capture processes (the s-process and r-process respectively).  The s-process signature of the solar system abundances can be split into three components; the weak-component, which involves nuclei with $60 < A < 90$, the main-component, which involves nuclei having atomic mass between $90 < A < 208$, and the strong component, which accounts for the production of the solar $^{208}$Pb \cite{1989RPPh...52..945K}.  The s-process site is attributed to massive stars for the weak component, and to AGB stars having initial mass between 1 and 3$M_{\odot}$ at solar-like metallicity for the main component and at low metallicity for the strong component \cite{1998ApJ...497..388G}.  Stellar evolution models of massive stars can be used to determine information on the conditions within stellar interiors and calculate the s-process yields relevant for the weak-component.  Any changes to the input physics, such as improved laboratory nuclear reaction rates, can affect the evolution of the simulated star and consequently affect the s-process yields.  Therefore, different nuclear reaction rates can be tested for their astrophysical impact.  In this work, we will show how variations in the rate of the $^{12}$C + $^{12}$C reaction affect the evolution of a 25$M_{\odot}$ star and the consequential s-process yields of the star.  The motivation for this work originates from nuclear physics experiments and theory (see for instance Spillane et al. 2007 \cite{2007PhRvL..98l2501S} and Gasques et al. 2007 \cite{2007PhRvC..76c5802G}), which are probing low enough energies to investigate the reaction within the Gamow window for $^{12}$C+$^{12}$C fusion.

\section{Carbon burning in massive stars and the s-process}

Carbon-core burning occurs at $T = 0.6 - 0.8$ GK.  The dominating reactions are $^{12}$C($^{12}$C,$\alpha)^{20}$Ne and $^{12}$C($^{12}$C,p)$^{23}$Na which, combined with the efficient $^{23}$Na(p,$\alpha)^{20}$Ne reaction, leaves $^{16}$O and $^{20}$Ne as the dominant isotopes in the core (the $^{16}$O being a remnant from the previous helium burning stage).  In addition to core burning, there are typically multiple stages of shell burning later in the evolution of the star, where the number of shells differs depending on the initial mass of the star.  The main neutron source for the s-process in massive stars is $^{22}$Ne by the reaction $^{22}$Ne$(\alpha,$n$)^{25}$Mg. $^{22}$Ne is produced in helium burning from the $^{14}$N$(\alpha,\gamma)^{18}$F$(\beta^{+})^{18}$O$(\alpha,\gamma)^{22}$Ne reaction chain during the supergiant phase.  $^{22}$Ne is partially burnt at the end of helium burning with a neutron density, $n_n \sim 10^6$ cm$^{-3}$ (see, for example, Raiteri et al. 1991 \cite{1991ApJ...367..228R}).  The $^{22}$Ne left in the CO-core is then burnt during carbon-shell burning with higher neutron densities ($n_n \sim 10^{11} - 10^{12}$cm$^{-3}$), with the $^{12}$C($^{12}$C,$\alpha)^{20}$Ne reaction providing the $\alpha$ particles \cite{2007ApJ...655.1058T}.  The carbon-shell burning is sensitive to the profile of $^{12}$C after core-carbon burning, which is in turn sensitive to the still uncertain $^{12}$C$(\alpha,\gamma)^{16}$O rate and the choice of convection physics in the models \cite{2007ApJ...655.1058T}\cite{2001ApJ...558..903I}\cite{2004ApJ...611..452E}.  In addition, the yields are sensitive to changes in reactions involving $^{22}$Ne or its formation \cite{2007ApJ...655.1058T}\cite{1994ApJ...437..396K}.  Most of the s-process-rich material ejected by the supernova event of a 25$M_{\odot}$ star is formed by carbon-shell burning, affecting the weak s-component \cite{1991ApJ...371..665R}\cite{2001ApJ...549.1085H}.

Stellar models of a 25$M_{\odot}$ star with metallicity $Z=0.01$ were generated using the Geneva Stellar Evolution Code \cite{2008Ap&SS.316...43E}, with the nuclear network post-processed using the NuGrid PPN tool \cite{2008nuco.confE..23H}.  The post-processing was computed using the KHAOS cluster at Keele University.  Variations in the carbon burning rate were chosen in relation to the `standard' Caughlan and Fowler rate (CF88 from now on).  Models were generated with the CF88 rate (referred to as model C12s in this work), the CF88 rate multiplied by 10 (C12t10) and divided by 10 (C12d10).  The ratio of the p- and $\alpha$-channels was chosen to be 35\%:65\% \cite{2006PhRvC..73f4601A}.  The additional uncertainty associated with this choice of ratio will be investigated in a forth-coming paper.

\begin{figure}[htb]
   \centering 
   \includegraphics[scale=0.5, bb=0 0 1292 576, width=\textwidth]{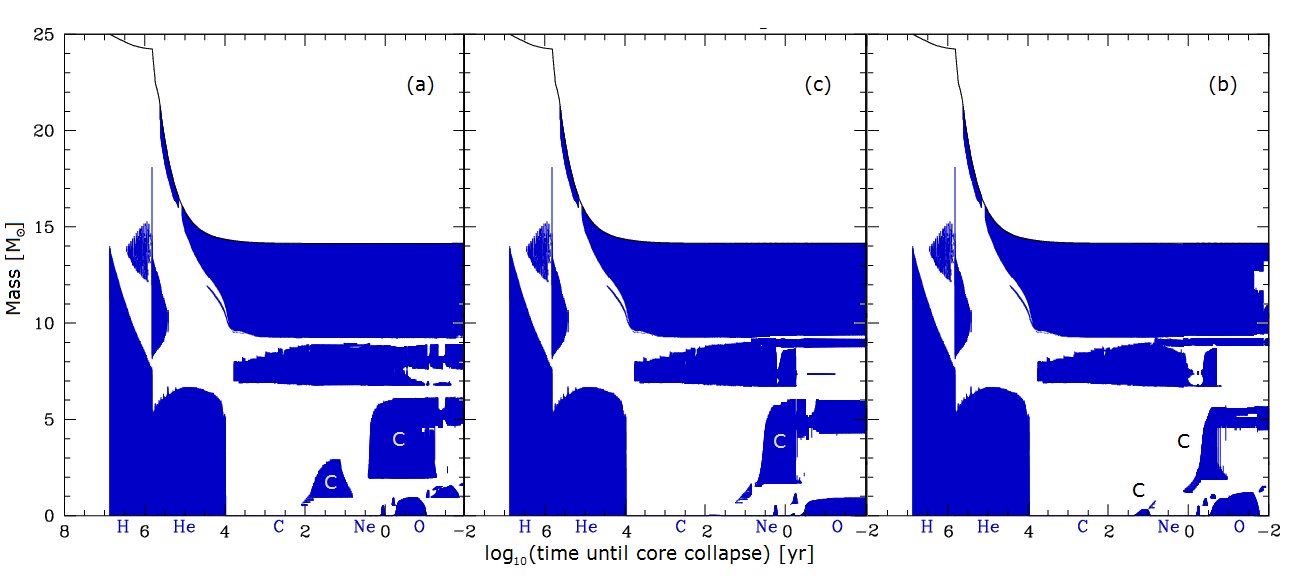}    
   \caption{Structure evolution diagrams of 25$M_{\odot}$ models at solar metallicity using different $^{12}$C burning rates.    Here, (a) is model C12t10, (b) is model C12s and (c) is model C12d10.  The shaded regions correspond to convection zones present in the star, with the type of burning indicated by labels.}
   \label{fig:kips}
\end{figure}

Differences in the convective history can be seen in each of the different models.  Fig. \ref{fig:kips} shows structure evolution diagrams for the models C12t10, C12s and C12d10.  In each case the convection zones for carbon shell burning at $\log_{10}$(time in years until core collapse) $\sim$ 1 to -2 differ in size and duration.  For model C12t10, the duration of the second carbon-shell burning process is much longer and the size of the convection zones is larger.  A large fraction of overlap between the first shell and the second shell is observed and isotopes produced in the first shell by the s-process will be mixed into the second.  The presence of overlapping carbon shells was previously noted by Chieffi et. al. \cite{1998ApJ...502..737C}.  For model C12d10, the second carbon-shell burning episode occurs later in time than for the other two models, which is due to the star contracting further before it can reach a temperature high enough so that carbon burning is activated in the shell.

\begin{figure}[htb]
   \centering 
   \includegraphics[scale=0.5, bb=0 0 1800 986, width=\textwidth]{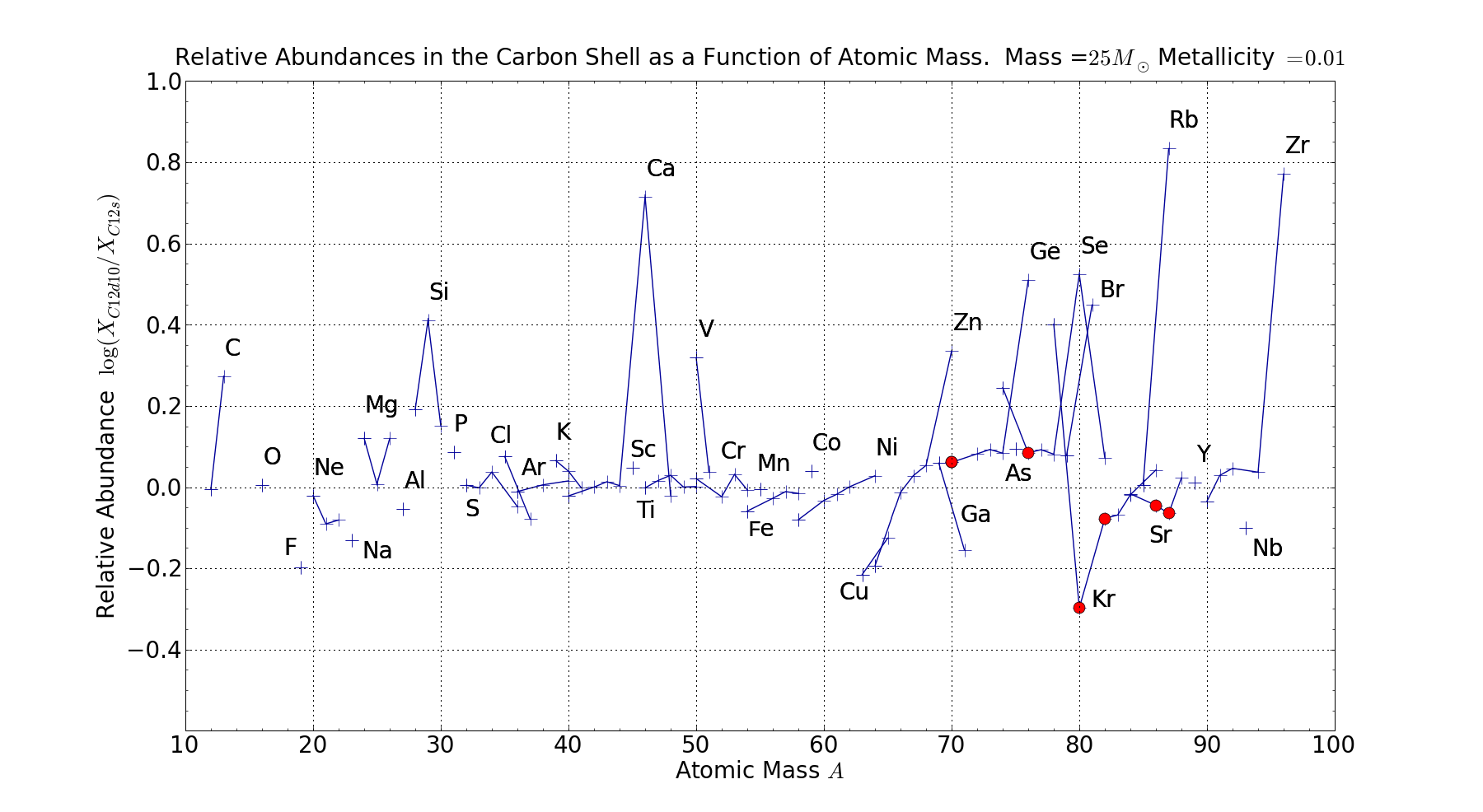}    
   \caption{The relative abundances of stable isotopes up to Niobium between models C12d10 and C12s.  S-only isotopes are indicated with a red circle.}
   \label{fig:ab_C12d10}
\end{figure}

\begin{figure}[htb]
   \centering 
   \includegraphics[scale=0.5, bb=0 0 1900 986, width=\textwidth]{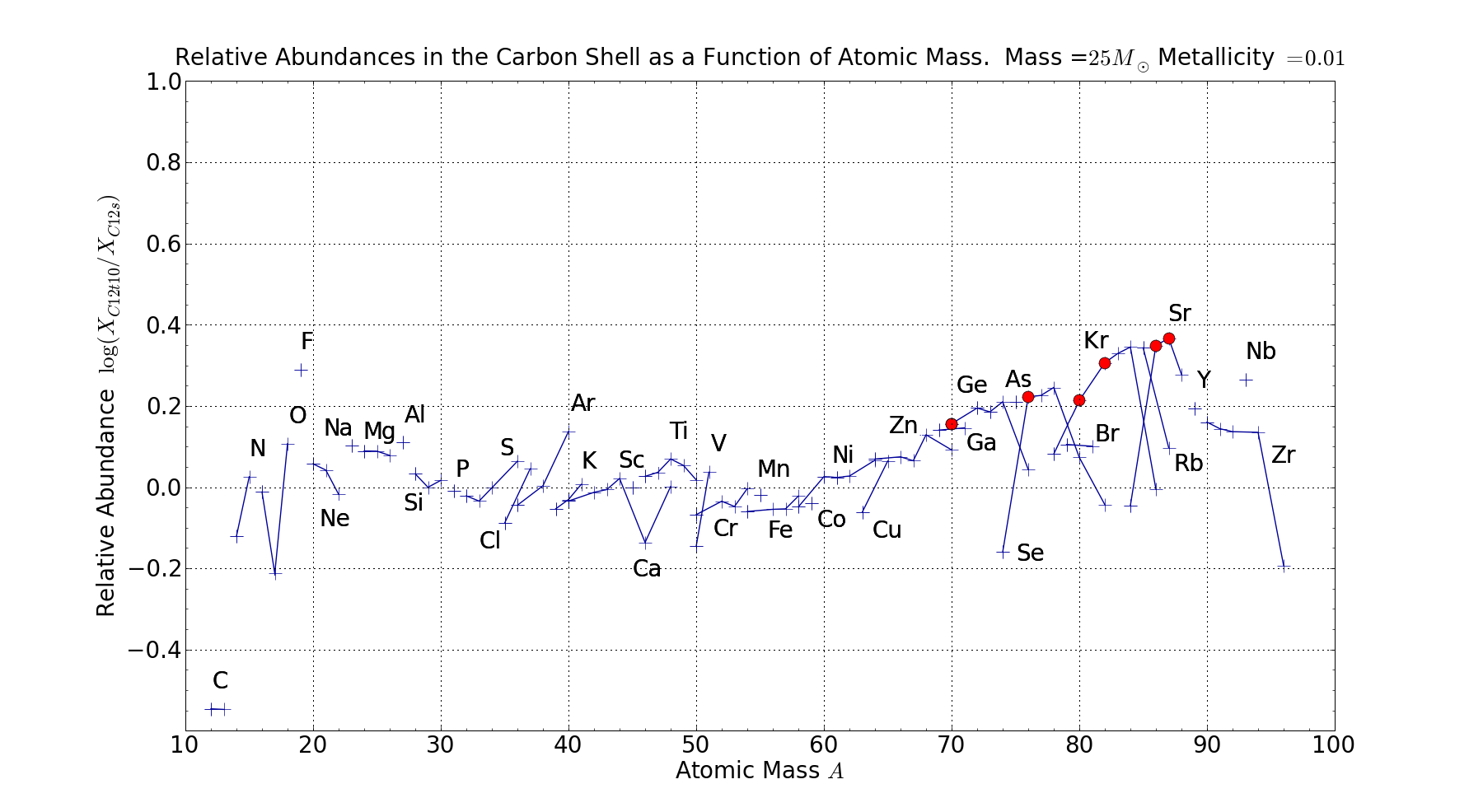}    
   \caption{The relative abundances of stable isotopes up to Niobium between models C12t10 and C12s.  S-only isotopes are indicated with a red circle.}
   \label{fig:ab_C12t10}
\end{figure}

Fig. \ref{fig:ab_C12d10} shows the relative abundances of stable isotopes in the second carbon shell of the C12d10 model with respect to the C12s model.  Fig. \ref{fig:ab_C12t10} shows the equivalent plot for the C12t10 model with respect to the C12s model.  In fig. \ref{fig:ab_C12d10}, a clear signature of a higher neutron density is shown, e.g. higher production of r-only species $^{70}$Zn, $^{76}$Ge and lower production of $^{80}$Sr due to the $^{79}$Se branching.  The reason is that the star in model C12d10 contracts further, resulting in carbon shells that burn at a higher temperature than the C12s model.  Therefore the $^{22}$Ne neutron source is burned at a higher temperature, increasing the neutron density.  The s-process efficiency is quite similar in model C12s and C12d10 (see, for example, the similar production of neutron magic isotope $^{88}$Sr).  Concerning model C12t10, in fig. \ref{fig:ab_C12t10}, a general increase in the s-process efficiency for isotopes with $60 < A < 90$ is shown, compared to model C12s.  The overlap between the first and second convective shells causes the initial distribution of isotopes at the start of the second carbon-shell burning to be affected by the products of the first shell.  In the first shell, neutrons are also efficiently produced by an additional neutron source, $^{17}$O($\alpha$,n)$^{21}$Ne, where $^{17}$O is mostly produced by $^{16}$O(p,$\gamma$)$^{17}$F($\beta^+$)$^{17}$O.

In this paper, it has been shown that changes to the carbon burning rate by a factor of 10 in stellar models of a 25$M_{\odot}$ star significantly affect the s-process yields.  The overlap of convective carbon shells active at different burning temperatures also has important implications for the s-process yields.  Further analysis will be conducted in a forth-coming paper.

\section*{References}

\providecommand{\newblock}{}

\end{document}